\documentclass[5p]{elsarticle}

\usepackage{amssymb}
\journal{Nuclear Instruments and Methods A}

\begin{document}

\begin{frontmatter}

%% Title, authors and addresses

%% use the tnoteref command within \title for footnotes;
%% use the tnotetext command for the associated footnote;
%% use the fnref command within \author or \address for footnotes;
%% use the fntext command for the associated footnote;
%% use the corref command within \author for corresponding author footnotes;
%% use the cortext command for the associated footnote;
%% use the ead command for the email address,
%% and the form \ead[url] for the home page:
%%
%% \title{Title\tnoteref{label1}}
%% \tnotetext[label1]{}
%% \author{Name\corref{cor1}\fnref{label2}}
%% \ead{email address}
%% \ead[url]{home page}
%% \fntext[label2]{}
%% \cortext[cor1]{}
%% \address{Address\fnref{label3}}
%% \fntext[label3]{}

\title{Application of Geiger-mode photo sensors in Cherenkov detectors}

%% use optional labels to link authors explicitly to addresses:
%% \author[label1,label2]{<author name>}
%% \address[label1]{<address>}
%% \address[label2]{<address>}

\author[A,B]{Ahmed Gamal\corref{D}}
\author[A]{B\"uhler Paul}
\author[A]{Cargnelli Michael}
\author[C]{Hohler Roland}
\author[A]{Marton Johann}
\author[C]{Orth Herbert}
\author[A]{Suzuki Ken}
\address[A]{Stefan Meyer Institute for Subatomic Physics of the Austrian Academy
of Sciences, Vienna, Austria}
\address[B]{Al-Azhar University, Faculty of Science, Physics Department, Cairo,
Egypt}
\address[C]{GSI Helmholtzzentrum f\"ur Schwerionenforschung GmbH, Darmstadt,
Germany}
\cortext[D]{Corresponding author: gamal.ahmed@assoc.oeaw.ac.at}

\begin{abstract}

Silicon-based photosensors (SiPMs) working in the Geiger-mode represent an
elegant solution for the readout of particle detectors working at low-light
levels like Cherenkov detectors. Especially the insensitivity to magnetic fields
makes this kind of sensors suitable for modern detector systems in subatomic
physics which are usually employing magnets for momentum resolution. In our
institute we are characterizing SiPMs of different manufacturers for selecting
sensors and finding optimum operating conditions for given applications.
Recently we designed and built a light concentrator prototype with 8$\times$8
cells to increase the active photon detection area of an 8$\times$8 SiPM
(Hamamatsu MPPC S10931-100P) array. Monte Carlo studies, measurements of the
collection efficiency, and tests with the MPPC were carried out. The status of
these developments are presented.

\end{abstract}

\begin{keyword}
%% keywords here, in the form: keyword \sep keyword

Silicon photomultiplier \sep Light concentrator \sep position-sensitive
photodetector matrix \sep Cherenkov detector

%% MSC codes here, in the form: \MSC code \sep code
%% or \MSC[2008] code \sep code (2000 is the default)

\end{keyword}

\end{frontmatter}

% ----------------------------------------------------------------------------
%%
%% Start line numbering here if you want
%%
% \linenumbers

%% main text
\section{Introduction}
\label{intro}

The position sensitive detection of Cherenkov light is extremely important for
particle identification in many particle physics experiments like in the
upcoming PANDA experiment \cite{PB09a} at FAIR. The challenge is the efficient
detection of photons at low light levels with very good timing performance. The
field of photon detection has experienced considerable progress with the
development of matrix-APDs operated in the Geiger-mode. One accepted name of
such a device is silicon photomultiplier (SiPM). SiPMs have high amplification
in the order of $10^6$ similar to traditional photomultipliers (PM). The main
advantages are the rather high photo detection efficiency of some devices in the
blue wavelength range (about twice compared to PMs), sub-nanosecond timing,
compact and robust design. The insensitivity to even high magnetic fields is a
major surplus in modern particle detectors which are usually employing magnetic
fields. Typical applications of SiPMs studied at the Stefan Meyer Institute are
the readout of scintillating fiber detectors \cite{suzuki09a} and timing
detectors using the fast Cherenkov process \cite{ahmed09a}. There are attempts
to construct arrays of SiPMs for RICH (see e.g. \cite{dolenec10a}). In order to
select the SiPM type with the best performance in photon detection efficiency,
dark count rate, and timing performance a study of the characteristics was
performed \cite{ahmed09a}. For position sensitive devices the small format
factor calls for the increase of the detection area with light concentrators
like Winston cones, small lenses, or mirror-type light concentrators. We are
reporting on the development, Monte Carlo simulation, and test of a 8$\times$8
cell light concentrator and the measured characteristics of the selected SiPM
type (Hamamatsu MPPC S10931-100P with 3$\times$3 mm$^2$ active area) which will
be arranged in an 8$\times$8 matrix to be readout with integrated electronics.

% ----------------------------------------------------------------------------
\section{Light concentrator}

The function of a light concentrator is to enhance the number of photons
reaching a detector above the number which is present without concentrator. In
contrast to an imaging system, image conservation is not required. Many
different and fairly complex systems have been proposed to optimize light
concentration in various applications. Having however an application in a large
high energy physics experiment in mind we considered a simple but robust and
easy to fabricate solution.

\begin{figure}[t] 
\centering 
\includegraphics[height=0.45\textwidth,keepaspectratio,angle=-90]{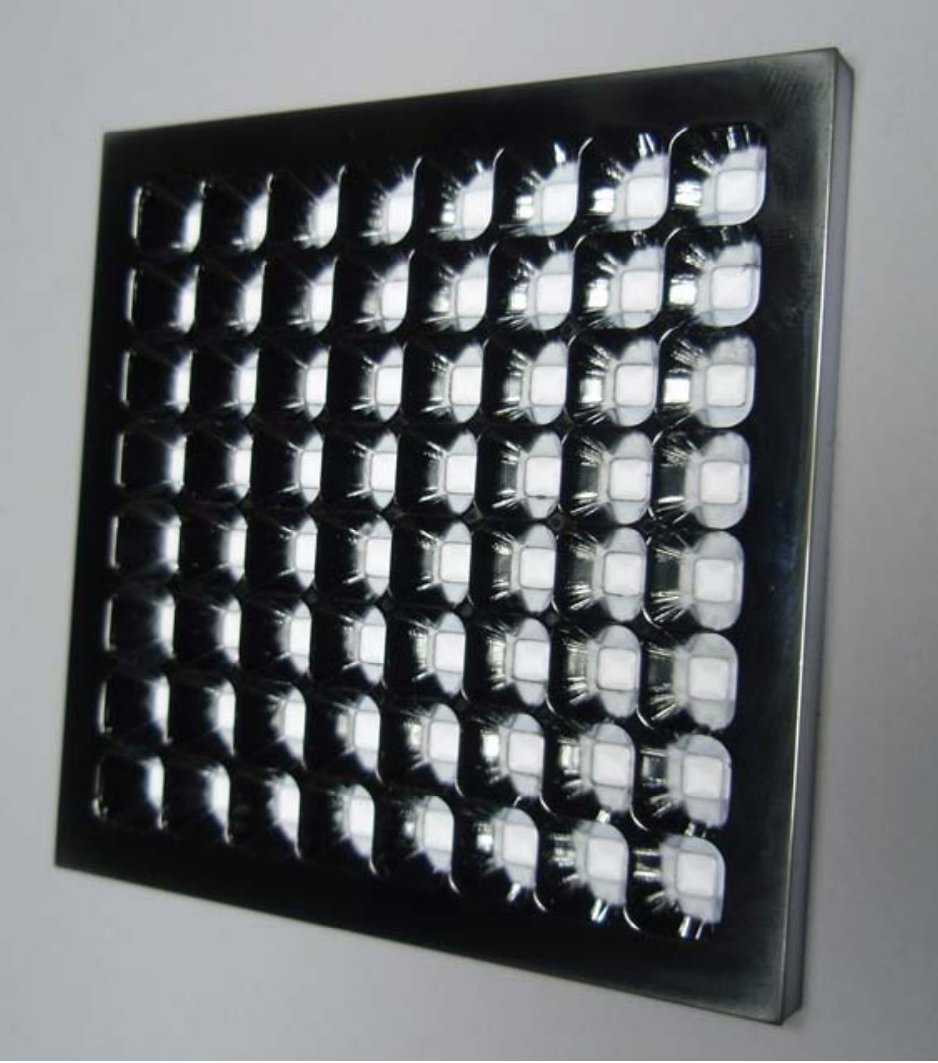}
\caption{Photograph of the light concentrator. The chrome-plated 4.5 mm thick
brass plate contains 64 pyramid-shaped funnels with quadratic (rounded edges)
entrance and exit apertures of 7$\times$7 mm$^2$ and 3$\times$3 mm$^2$, respectively. }
\label{fig01}
\end{figure}

The light concentrator consists of a plate of 8.4$\times$8.4 cm$^2$ containing
64 regularly arranged pyramid-shaped funnels with quadratic (rounded edges)
entrance and exit apertures of 7$\times$7 mm$^2$ and 3$\times$3 mm$^2$,
respectively. It was designed to be combined with a SiPM array consisting of
8$\times$8 MPPCs with 3$\times$3 mm$^2$ active area. The concentrator is made
out of brass and the funnels were produced by electroerosion. Afterward the
plate was chrome-plated. A photograph is displayed in figure \ref{fig01}. The
height of the concentrator plate (length of funnels) used for this study is 4.5
mm.

% ............................................................................
\subsection{Monte Carlo Studies of the light concentration efficiency}

\begin{figure}[t] 
\centering 
\includegraphics[width=0.48\textwidth,keepaspectratio]{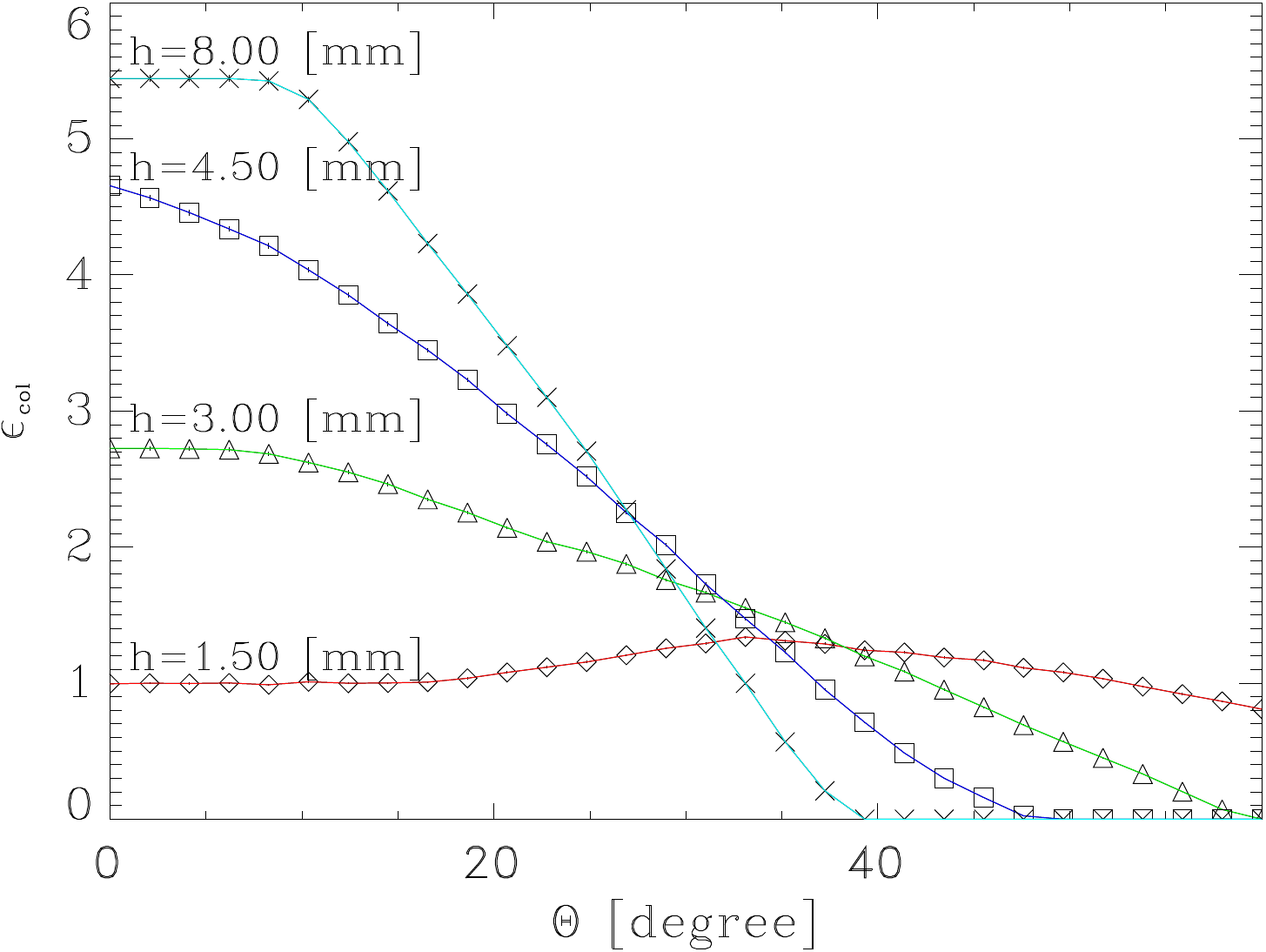}
\caption{Light collection efficiency as function of angle of incidence $\Theta$
(azimuthal angle of incidence is 0) for different values of the funnel length
$h$. Specular reflection is assumed. The entrance points of the photons to the
funnels was distributed homogeneously in the entrance aperture.}
\label{fig02}
\end{figure}

In order to estimate the collection efficiency of the light concentrator and to
study its dependence on the length of the funnels and the angle of incidence of
the photons we carried out Monte Carlo simulations. Photons with given direction
were produced at the entrance aperture and their path was followed until they
were either absorbed by the funnel walls or left the funnel through either of
the apertures. The collection efficiency $\epsilon_{col}$ of a single funnel is
defined by

\begin{equation}
\epsilon_{col} = \frac{n_d/N_{phot}}{(3/7)^2}
\label{equ01}
\end{equation}

where $n_d$ is the number of photons reaching the exit aperture and $N_{phot}$
is the total number of simulated photons penetrating the entrance aperture
(typically $\approx 10^5$). If all photons entering the entrance aperture of a
funnel were transported to the exit aperture a collection efficiency of
$(7/3)^2=5.4$ would result. The optical properties of the funnel walls were
defined by a reflection coefficient and a factor characterizing the roughness of
the surfaces. In figure \ref{fig02} results are shown for the idealized case of
specular reflection (neglecting absorption and assuming perfectly smooth
surfaces). The points where the photons entered the funnel were distributed
homogeneously over the entrance aperture. The such determined collection
efficiency is drawn as function of angle of incidence $\Theta$ (angle relative
to the aperture normal) for different values of the funnel length $h$. In these
simulations the azimuthal angle of the incident photons was kept constant and
set to 0 degrees.

At normal incidence ($\Theta=0$ degrees) the collection efficiency increases
with increasing funnel length and reaches the optimum value of 5.4 at $h>6$ cm.
At larger angles ($\Theta>30$ degrees) the collection efficiency is generally
smaller than at normal incidence. However, here the relation between
$\epsilon_{coll}$ and $h$ changes, such that the collection efficiency is
largest at small funnel length and decreases with increasing funnel length.
Note, that since in these simulations the absorption of the photons in the
funnel walls is neglected the only way for photons to escape from the funnel
without reaching the exit aperture is to leave upwards through the entrance
aperture. This happens to photons which enter the funnel close to a corner and
are reflected from funnel wall to funnel wall and finally back to the entrance
aperture.

The optimum choice of the height of this kind of light concentrator depends on
the expected angular distribution of the incident light. To estimate the
experimentally achievable collection efficiency, the true spatial and angular
distribution of the incident photons must be known and the absorption of the
photons on the walls of the funnels and also diffusive reflection need to be
taken into account.

The effect of the absorption can be roughly estimated without repeating the
simulations with modified optical properties of the funnel walls by multiplying
the results shown in figure \ref{fig02} with a factor $f_{abs}^{n_{ref}}$, where
$f_{abs}$ is the absorption coefficient of the funnel wall and $n_{ref}$ is the
number of reflections the photon experiences before it exits through the exit
aperture. $n_{ref}$ is a function of $h$, the angle of incidence, and the
position the photon enters the funnel. In the simulated situation, the average
number of reflections at a given \{$h$, $\Theta$\} pair is largest at large $h$
and small $\Theta$ and smallest at small $h$ and large $\Theta$. Thus for a
given light source, a geometry with large $h$, which seems best in the
simulations without absorption, might turn out to be less efficient than a
geometry with smaller $h$, when taking absorption into account.  The light
concentrator we are experimenting with has a height of 4.5 mm. If we assume a
reflection coefficient of 0.55 (Chromium at 400 nm \cite{refind}) we find from
the simulations an average light collection efficiency of around 2.8 at normal
incidence, which falls below 1 at $\Theta>30$ degrees (with an average value of
1.5 for $0<\Theta<60$ degrees, case of uniformly distributed angles of
incidence).

% ............................................................................
\subsection{Measurement of the light concentration efficiency}

In order to experimentally test the collection efficiency of the light collector
we used two different settings.

\begin{figure}[ht] 
\centering 
\includegraphics[height=0.41\textwidth,keepaspectratio,angle=90]{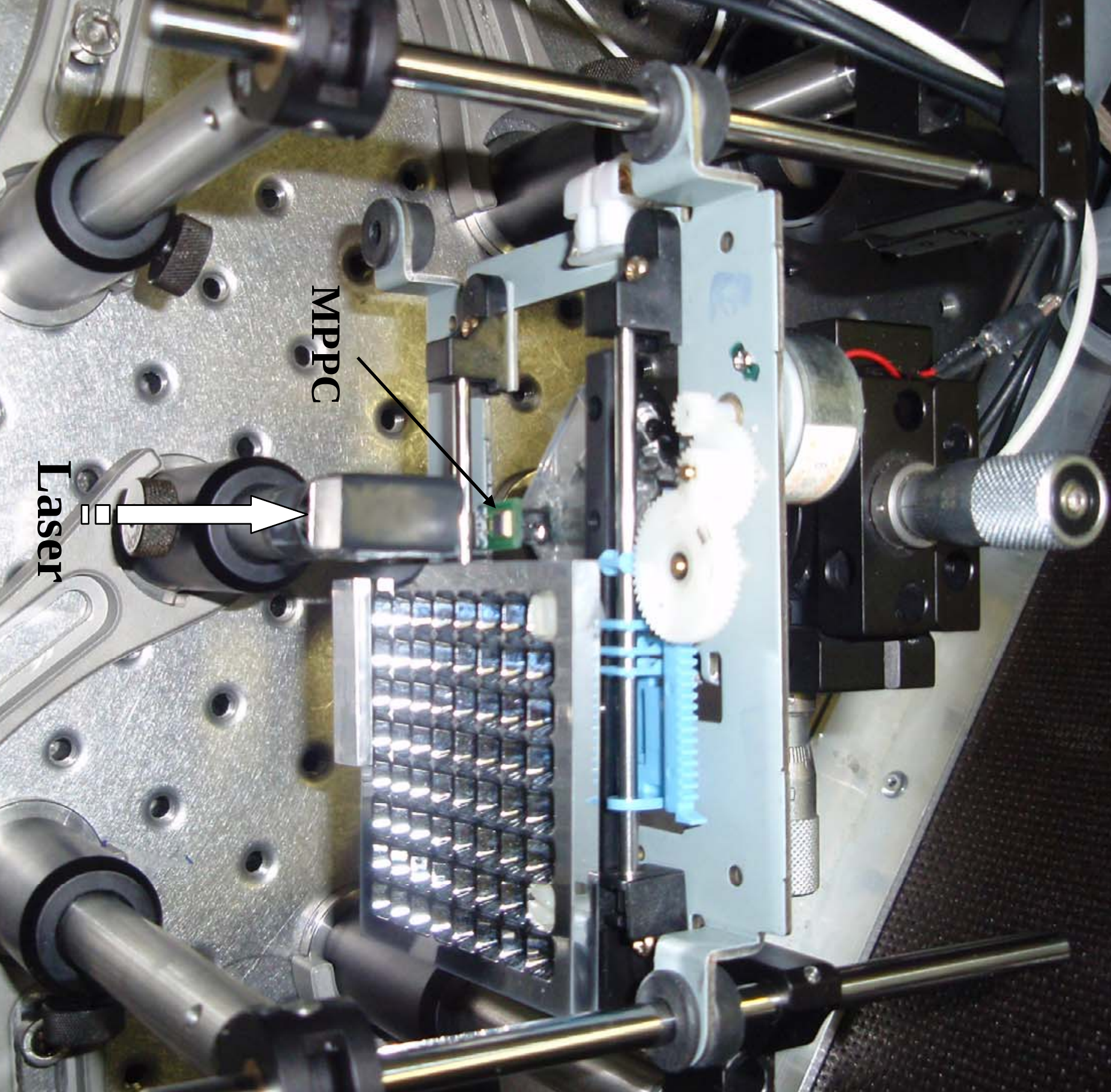}
\caption{Photograph of the setup used to measure the collection efficiency of
the light concentrator. A plexiglas bar which is illuminated with blue laser
light serves as light source. An MPPC was used to measure the light intensity.
The light concentrator is mounted on a sliding bar between light source and
photo sensor.}
\label{fig03}
\end{figure}

The first setup is shown in figure \ref{fig03}. The light source was the squared
end face of a plexiglas bar, which was illuminated from the other side with
diffused light from a blue laser (408 nm). Along the long sides the plexiglas
bar was wrapped into aluminum to enhance the internal reflection. With this a
fairly homogeneous light source was provided with a presumably broad (although
not well defined) angular distribution of the emitted photons. As photo sensor
we used a 3$\times$3 mm$^2$ MPPC (S10931) sensor from Hamamatsu. It was mounted
opposite to the end face of the plexiglas bar in a distance of $\approx$5 mm
with the sensitive surface facing the light source. The light concentrator was
mounted on a sliding bar, between light source and MPPC. This allowed to move
the light concentrator, remotely controlled, in and out of the path of the light
without altering the relative position of light source and photo sensor. The
in-position of the light concentrator was adjusted such that the exit aperture
of one funnel exactly matched the sensitive area of the MPPC. The entire setup
was mounted in a black box.

In a series of measurements we determined the signal height registered by the
MPPC with the light concentrator in in- ($s_{in}$) and out-position ($s_{out}$)
for different funnels. The collection efficiency $\epsilon_{col}$ is computed
with $\epsilon_{col}=s_{in}/s_{out}$. We found an average value of
$\epsilon_{col}\approx 1.8\pm 0.2$.

This is close to the value of 1.5 which is found by simulations for the case of
homogeneously spatial distributed light and uniform distribution of the angle
of incidence between 0 and 60 degrees assuming an reflection coefficient of the
funnel walls of 0.55.

In the second setup, the light concentrator was irradiated with nearly parallel
and normal incident light (estimated maximum $\Theta$-angle of 4 degrees). As
photosensor a photodiode of 10$\times$20 mm$^2$ active area was used. The light
concentrator was covered with a blind leaving two funnel apertures open. The
light throughput was measured once with the light illuminating the top face of
the light concentrator ($s_{top}$) and once from the bottom ($s_{bottom}$).
$\epsilon_{col}$ is then obtained by the ratio of the two measurements
$\epsilon_{col}=s_{top}/s_{bottom}$. For this setup we found
$\epsilon_{col}\approx 2.5$. This is comparable with the factor 2.8 which
results from simulations of normal incident light, assuming a reflection
coefficient of 0.55 of the funnel walls. The results seem promising and justify
further investigations.

% ----------------------------------------------------------------------------
\section{Characterization of SiPMs for the readout array}

As readout device for the light concentrator we consider to use the recent large
active area SiPMs - the Multi-Pixel Photon Counters (MPPCs) from Hamamatsu
\cite{hamamatsu}. These MPPC photo sensors are sensitive in the blue-light range
($\approx$ 400 nm) which matches well with the light emitted by scintillators
and Cherenkov radiators. From the series of MPPCs with an active area of
3$\times$3 mm$^2$ we selected the type S10931-100P because of its large fill
factor (78.5\%) and the very small housing which allows for a compact
arrangement of several devices in a matrix for readout of the entire light
collector.

Besides the fact, that SiPM are resistant against magnetic fields, the most
relevant performance figures of MPPCs in subatomic-physics applications is the
dark count rate and the time resolution. Especially in cases where only few
photons carry the physical information the ability to detect single photons is
very appealing. This intrinsic ability of SiPMs can however be limited by large
dark count rates. Fast timing e.g. allows for higher processing rates and more
precise determination of physical quantities \cite{schepers09a}.

In view of a possible usage of these photo sensor for such applications we
measured its dark count rate and timing performance as function of the
temperature and over-voltage (difference between the applied bias voltage and
the breakdown voltage). Here we present preliminary measurements taken at a
temperature of $-10^{\circ}$C.

\subsection{Dark count (noise) rate}

\begin{figure}[t] 
\centering 
\includegraphics[width=0.48\textwidth,keepaspectratio]{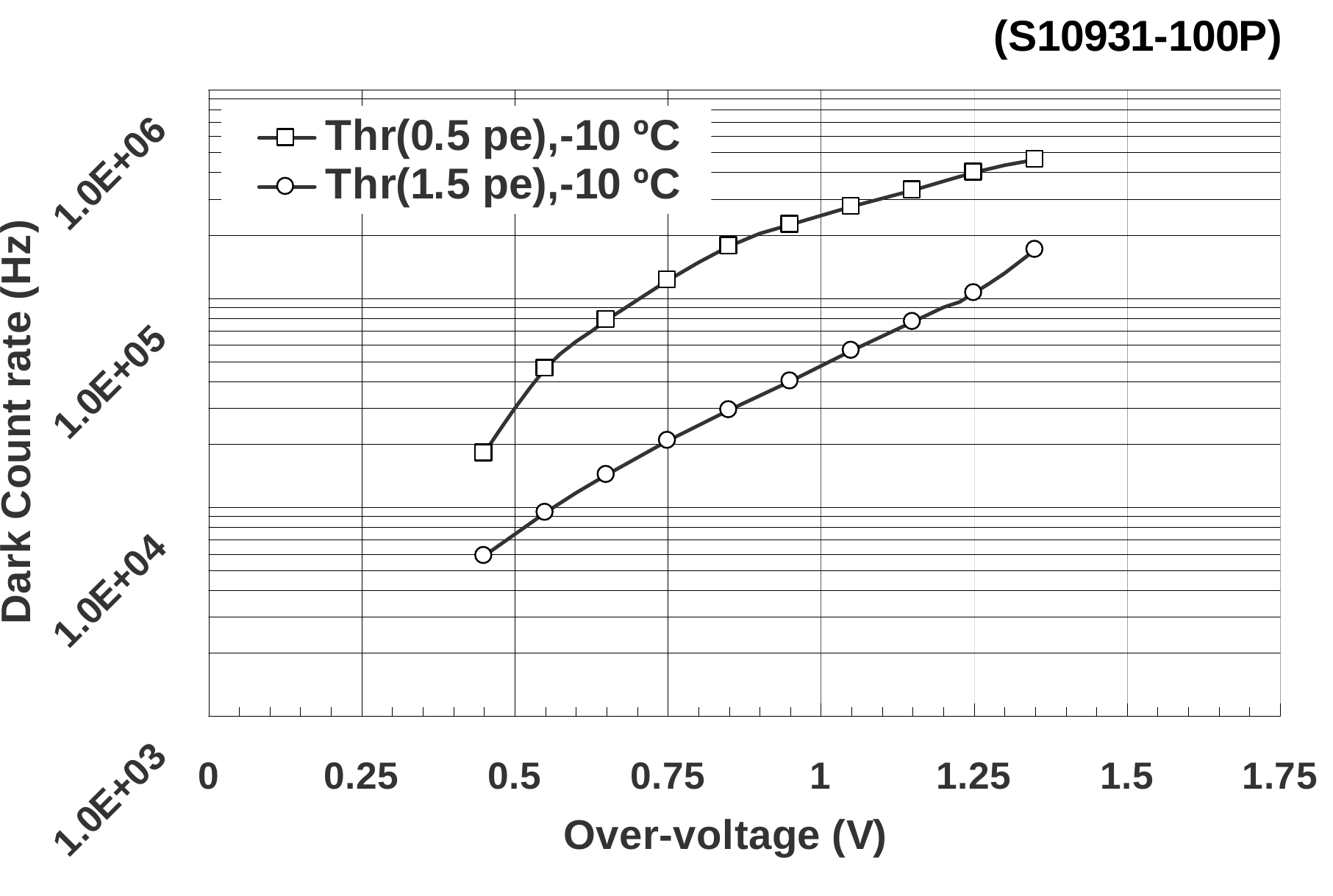}
\caption{Dark count rate as a function of over-voltage, at $-10^{\circ}$C.
Square and circular symbols correspond to data points at 0.5~p.e. and 1.5~p.e.
thresholds levels, respectively.}
\label{fig04}
\end{figure}

The MPPCs are solid-state devices which generate noise due to thermal excitation
\cite{hamamatsu}. The noise rate is measured by counting the number of output
signals with an amplitude exceeding a given threshold value in absence of an
external light input. Figure \ref{fig04} shows the measured dark count rates at
$-10^{\circ}$C and at different over-voltage values. The threshold level was
adjusted to be 0.5 and 1.5 photoelectrons (p.e.). 

At a threshold of 0.5~p.e. the dark count rate increases from around 20~kHz at
an over-voltage of 0.5~V to approximately 0.5~MHz at 1.3~V over-voltage. The
dark count rate drops rapidly with increasing threshold levels. By increasing
the threshold to 1.5~p.e. the dark count rate is reduced by approximately a
factor 3.

\subsection{Timing performance}

The time resolution of a SiPM depends on the operation conditions and the number
of detected photons. Especially in applications with low light levels a proper
selection of the operation temperature and voltage is essential to obtain good
timing performance \cite{ahmed10a}. In order to measure the time resolution of
the S10931-100P MPPC the device was stimulated with short light pulses from a
blue laser. Amplitude and time of the response was recorded and the time
resolution was determined by the standard deviation of the recorded response
time. In order to obtain the time resolution as function of number of fired
pixels (NFP) the measured amplitude distribution was split into narrow bins,
corresponding to 1-2 fired pixels, and the time resolution was determined
separately for each bin. A detailed description of the experimental setup and
methods is given in \cite{ahmed10a}.

\begin{figure}[t] 
\centering 
\includegraphics[width=0.48\textwidth,keepaspectratio]{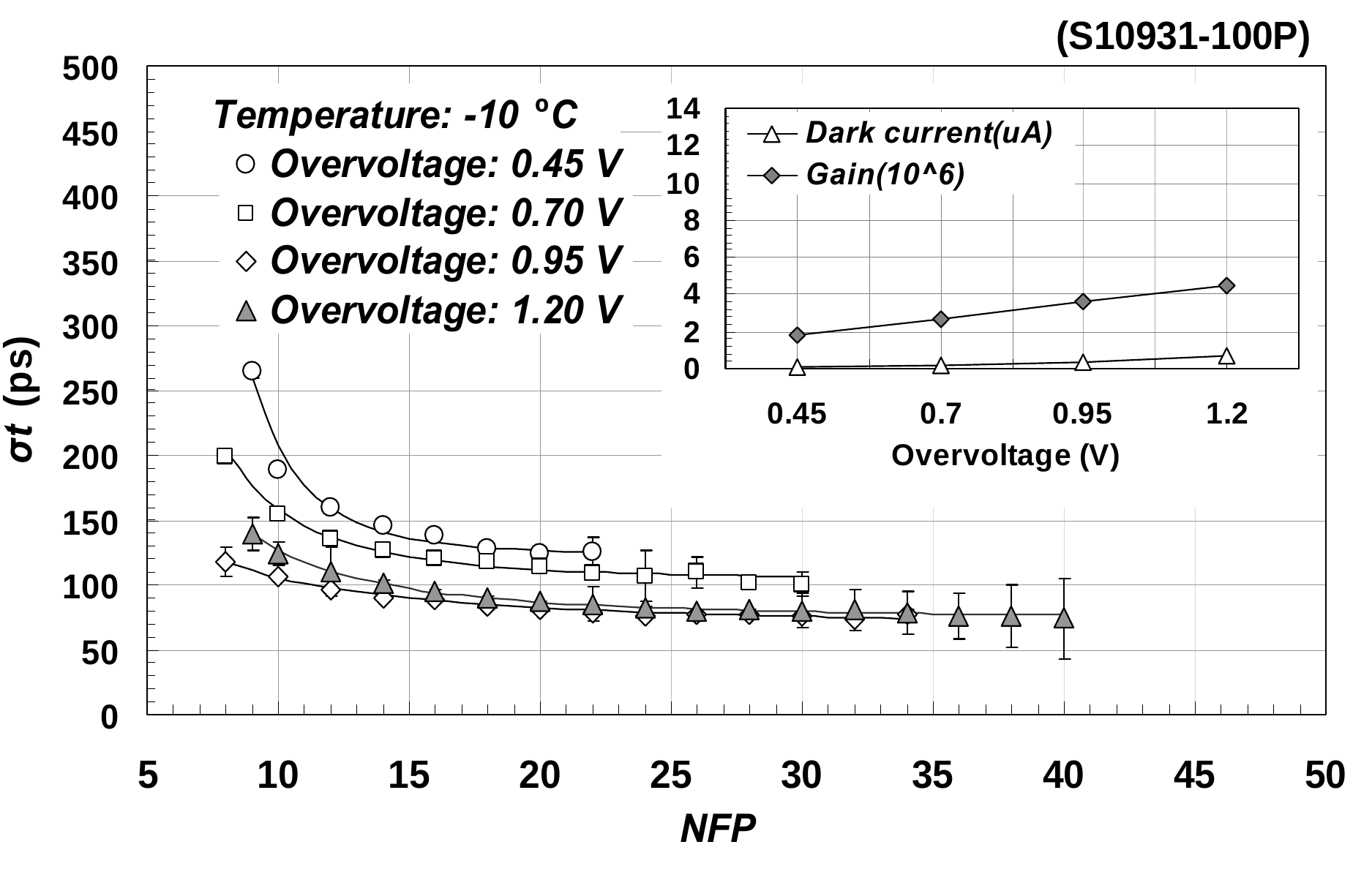}
\caption{Time resolution as function of number of fired pixels (NFP) of the MPPC
S10931-100P at $-10^{\circ}$C, and with different over-voltages. The insets show
gain and dark current as function of over-voltage.}
\label{fig06}
\end{figure}

The results of these measurements are summarized in figure \ref{fig06}. The time
resolution is plotted as NFP for different values of over-voltage at a
temperature of $-10^{\circ}$C. In the insets the gain and dark current are
displayed as function of over-voltage.

% \begin{figure}[hbt] 
% \centering 
% \includegraphics[height=0.5\textwidth,keepaspectratio,angle=90]{RICHproceedings_fig02.pdf}
% \caption{Charge distributions of the MPPC (S10931-100P) at $-10^{\circ}$C and different
% over-voltages. The inset shows the distribution acquired at an over-voltage of
% 0.75 V but at lower light intensity than the distribution below. In this case
% single photon peaks are resolved which can be explored to determine the gain of
% the device at the given operation condition.}
% \label{fig05}
% \end{figure}

The dependence of the time resolution on the over-voltage and NFP is similar to
what we found for the S10362-33-100C MPPC \cite{ahmed10a}. Especially
at small NFP an appropriate selection of the operation conditions is necessary
to obtain good timing performance. 

% ----------------------------------------------------------------------------
\section{Outlook}

The characterization of the light concentrator and photo sensor as presented
here, is the first step in a program which aims at the development of an
efficient and compact detector for the readout of RICH and DIRC detectors.
Further work is needed to study and optimize the collection efficiency (e.g.
replace chrome-plating by plating with better reflectivity) of the light
concentrator. Integration of 64 MPPCs into a matrix with the corresponding
front-end electronics is an additional task.

% ----------------------------------------------------------------------------
\section{Acknowledgments}

This work is partly supported by INTAS (project 05-1000008-8114) and
Hadronphysics2 (project 227431). One of us (G.A.) acknowledges the support by
the Egyptian Ministry of higher education.

%% The Appendices part is started with the command \appendix;
%% appendix sections are then done as normal sections
%% \appendix

%% \section{}
%% \label{}

% ----------------------------------------------------------------------------
%% The Appendices part is started with the command \appendix;
%% appendix sections are then done as normal sections
%% \appendix

%% \section{}
%% \label{}

%% References
%%
%% Following citation commands can be used in the body text:
%% Usage of \cite is as follows:
%%   \cite{key}         ==>>  [#]
%%   \cite[chap. 2]{key} ==>> [#, chap. 2]
%%

% ----------------------------------------------------------------------------
%% References with bibTeX database:

\bibliographystyle{elsarticle-num}
\bibliography{<your-bib-database>}

\begin{thebibliography}{00}

\bibitem{PB09a} PANDA Physics performance report, Strong Interaction Studies
with Antiprotons, PANDA Collaboration, arXiv:0903.3905v1 [hep-ex].
\bibitem{schepers09a} G.~Schepers et al., RICH for PANDA, Nucl. Instrum. Meth. A
598 (2009) 143.
\bibitem{suzuki09a} K.~Suzuki et al., Nucl. Instr. and Meth. A 610 (2009) 75.
\bibitem{ahmed10a} G.S.M.~Ahmed et al., Nucl. Instr. and Meth. A (in press),
arXiv:1006.4032 [physics.ins-det].
\bibitem{dolenec10a} R.~Dolenec et al., Nucl. Instr. and Meth. A (in press).
\bibitem{ahmed09a} G.S.M. Ahmed et al., J.Inst. 4 (2009) P09004.
\bibitem{refind} Refractive Index Database, http://refractiveindex.info/
\bibitem{hamamatsu}
http://jp.hamamatsu.com/products/sensor-ssd/4010/index\_en.htmli.
\bibitem{alds} Advanced Laser Diode Systems, http://www.alsgmbh.com/
\bibitem{photonique} Photonique SA, Switzerland (http://www.photonique.ch).

%% \bibitem{buzhan03a} P. Buzhan et al., Nucl. Instr. and Meth. A 504 (2003) 48.
%% \bibitem{renker06a} D. Renker, Nucl. Instr. and Meth. A 567 (2006) 48.
%% \bibitem{vinke09a} R.~Vinke et al., Nucl. Instr. and Meth. A 610 (2009) 188.
%% \bibitem{ronzhin10a} A. Ronzhin et al., Nucl. Instr. and Meth. A 616 (2010) 38-44.
%% \bibitem{collazuol10a} G.~Collazuol et al., Nucl. Instr. and Meth. A (in press).
%% \bibitem{stewart06a} A.G. Stewart et al., in: SPIE: Semiconductor Photodetectors III, vol. 6119, 2006.
%% \bibitem{mcclish07a} M. McClish et al., Nucl. Instr. and Meth. A 572 (2007) 1065.
%% \bibitem{deiters00a} K. Deiters et al., Nucl. Instr. and Meth. A 453 (2000) 223.
%% \bibitem{sadygov03a} Z. Ya. Sadygov et al., Nucl. Instr. and Methods A 504 (2003) 301.
%% \bibitem{golovin04a} V. Golovin and V. Saveliev, Nucl. Instr. and Methods A 518 (2004) 560.
%% \bibitem{zecotek} http://www.zecotek.com/

\end{thebibliography}

%% Authors are advised to submit their bibtex database files. They are
%% requested to list a bibtex style file in the manuscript if they do
%% not want to use elsarticle-num.bst.

%% References without bibTeX database:

% ----------------------------------------------------------------------------
\end{document}